\documentstyle[a4,cite,12pt,aps,epsfig]{revtex}

\def\nnd{\end{document}}
\def\dgr{\dagger}

\def\nnb{\nonumber}

\def\be{\begin{equation}}
\def\ee{\end{equation}}
\def\mn{\mu\nu}
\newcommand{\bea}{\begin{eqnarray}}
\newcommand{\eea}{\end{eqnarray}}

\def\sbra#1{\Big   ( #1  \Big   ) }
\def\mbra#1{\bigg  [ #1  \bigg  ] }
\def\bbra#1{\Bigg \{ #1  \Bigg \} }

\def\cma{\,,}
\def\hA{\widehat A}

\def\hWp{\widehat W^{+}}
\def\bWp{\overline W^{+}}
\def\hWm{\widehat W^{-}}
\def\bWm{\overline W^{-}}
\def\hZ{\widehat Z}
\def\bZ{\overline Z}

\def\wsep{ \nnb \\ &&}

\def\eed{\end{document}}
\def\ch{\overline h}
\def\lvec#1{{\stackrel{\leftharpoonup}{#1}}}
\def\rvec#1{{\stackrel{\rightharpoonup}{#1}}}

\def\be{\beta}

\makeatother

\begin{document}

\draft

\title{The one loop renormalization of
the effective Higgs sector and its implications}

\author{
             Qi-Shu YAN\footnote{
        E-mail Address: yanqs@mail.ihep.ac.cn} and
		Dong-Sheng Du\footnote{
        E-mail Address: duds@mail.ihep.ac.cn}\\
	Theory division,
        Institute of high energy physics,
	Chinese academy of sciences, Beijing 100039,
	Peoples' Republic of China
}
\bigskip

\address{\hfill{}}

\maketitle


\begin{abstract}
We study the one-loop renormalization the
standard model with anomalous Higgs couplings ($O(p^2)$)
by using the background field method,
and provide the whole divergence structure
at one loop level. The one-loop divergence
structure indicates that, under the quantum corrections,
only after taking into account
the mass terms of Z bosons ($O(p^2)$) and the whole bosonic
sector of the electroweak chiral Lagrangian ($O(p^4)$),
can the effective Lagrangian be complete up to $O(p^4)$.
\end{abstract}


\section{Introduction}
In the last paper \cite{our1}, we have considered the one-loop renormalization
of the electroweak chiral Lagrangian (EWCL) without Higgs boson
up to $O(p^4)$ and have derived its renormalization group equations.
The real world might exist a light Higgs boson, as preferred by
the supersymmetric models and indicated at the LEP.
So to take into account a light Higgs boson in the EWCL $O(p^4)$
is necessary in phenomenologies.

In this paper, we study the one loop divergence structures of
the effective Higgs sector up to $O(p^2)$ \cite{ewhiggs},
so as to ascertain the set of
complete operators up to $O(p^4)$. We also intend to
provide a reference to the renormalization of
the extended EWCL with effective Higgs sector.

The parameterization of the effective Higgs sector has
been conducted in the work of R. S. Chivukula and
V. Koulovassilopoulos \cite{ewhiggs}. The
phenomenologies of those anomalous couplings of Higgs sector has
been considered in \cite{hjhe}. Below just for the sake of
convenience, we formulate the effective Higgs sector as
\bea
\label{eq:sm}
{ \cal L} &=& - H_1 - H_2 - k' {(v_1+h)^2 \over 4} tr[D U^{\dagger} \cdot D U]
- { v_2^2 \over 4} tr[D U^{\dagger} \cdot D U]
\wsep - {1\over 2} \partial h \cdot \partial h
-{m_H^2 \over 2 } h^2 - {\lambda_3 \over 3!} h^3
- {\lambda_4 \over 4!} h^4 \,.\\
H_1&=&{1\over 4}  W_{\mn}^a W^{a \mn}\cma\\
H_2&=&{1\over 4}  B_{\mn}   B^{\mn}\cma
\eea
This way of parameterization is equivalent to that one given in
\cite{ewhiggs}. and the relations between our parameterization
and the one in \cite{ewhiggs} is
\bea
v_1 = V_0 {k \over k'},\,\, v_2 = V_0 \sqrt{1 - {k^2 \over k'}}\,,\,\,V_0^2=k' v_1^2 + v_2^2\,.
\eea
Where $v_1$ just indicates that part of the vacuum expectation value yields
by the Higgs potential of $h$, while the $v_2$ parameterize the extra
origin of vacuum expectation value different from the contribution of
$\langle h \rangle$.

The convention of this paper is as the same as given in
\cite{our1}, and we will conduct all the calculations
in the Euclidean space. The related definitions omitted here
can be found in \cite{our1}.

The structure of this paper is organized as following.
In the section II, we will introduce the background field gauge.
In the section III, we
provide the quadratic terms relevant to the
master formula of the one loop counter term. In the section IV,
we provide the necessary steps as how to extract the relevant
divergences. In the section V, we use the Schwinger proper time \cite{htkl}
and short distance expansion \cite{wkb} to extract the desired
divergences. We end this paper with some discussions and conclusions.

\section{The background field method and the gauge fixing terms}
In the spirit of the background field gauge quantization \cite{bfm},
we can decompose the Goldstone field into the classic part ${\overline U}$
and quantum part $\xi$
as
\bea
U\rightarrow {\overline U} {\widehat U}\,,\,\,
{\widehat U} = \exp\{{i 2 \xi \over \sqrt{k'} V }\}\,,
\label{upara}
\eea
where $V =v_1 + \ch$.
To parameterize the quantum Goldstone field in the above form is to simplify
the presentation of the standard form of quadratic terms.
The vector fields in the mass eigenstates are split as
\bea
V_{\mu}\rightarrow {\overline V_{\mu}} + {\widehat V_{\mu}}\cma
\eea
where ${\overline V_{\mu}}$ represents the classic background
vector fields and ${\widehat V_{\mu}}$ represents the quantum
vector fields.

By using the Stueckelberg transformation \cite{stuck} for the
background vector fields, we have
\bea
{\overline W}^{s\, a} \rightarrow U^{\dagger} {\overline W} U + i {\overline U}^{\dgr} \partial{\overline U} \cma
{\overline B} \rightarrow {\overline B}\,.
{\widehat W}^{s\, a} \rightarrow U^{\dagger} {\widehat W} U \cma
{\widehat B} \rightarrow {\widehat B}\cma
\eea
so the background Goldstone fields can completely be
absorbed by redefining the background vector fields,
and will not appear in the one-loop effective Lagrangian.

The Stueckelberg fields is invariant under the
gauge transformation of the background gauge fields,
such a property guarantees that the following computation
is gauge invariant with respect to the background
gauge transformation from the beginning if we can express
all effective vertices into the Stueckelberg fields.
After the loop calculation, by using the inverse
Stueckelberg transformation, the Lagrangian can be
restored to the form represented by its low energy
degrees of freedom.

Similarly, the Higgs scalar is split as
\bea
h = {\overline h} + {\widehat h}\,,
\eea
where the ${\overline h}$ and ${\widehat h}$ represent
the background and quantum Higgs, respectively.

The equation of motion of the background vector fields is given as
\bea
D_{\mu} {\cal \widetilde W}^{\nu\mu} = - \sigma_{0,V } V^{\nu}\,,\\
\sigma_{0,V }=k' R_2 {1 \over 4}  dia\{0,
G^2, g^2, g^2\} V ^2 .
\eea
with ${\cal \widetilde W}^{\mu\nu,T}=\{ A^{\mu\nu} + i e F_Z^{\mu\nu},
Z^{\mn} - i {g^2 \over G} F_Z^{\mn},
{\widetilde W^{+,\mn}},{\widetilde W^{-,\mn}} \}$.
The $\sigma_{0,V }$ is the mass matrix of the vector
boson. The EOM of vector bosons
derives the following relations
\bea
\partial \ln V  \cdot Z &=& - { R_2\over 2} \partial \cdot Z\,,\\
\partial \ln V  \cdot W^{+} &=& - {R_2 \over 2} \mbra{d \cdot W^{+} + i {1 \over 2} {g'^2 \over G} Z \cdot W^{+}}\,,\\
\partial \ln V  \cdot W^{-} &=& - {R_2 \over 2} \mbra{d \cdot W^{+} - i {1\over 2} {g'^2 \over G} Z \cdot W^{+}}\,,\\
R_2&=&{ k' \ch^2 + 2 V_0 k \ch + V_0^2 \over k' V^2}=1 + {v_2^2 \over k' V ^2}\,.
\eea
These relations are helpful to eliminate
terms, like $\partial \ch \cdot Z$, in the loop calculations.

The equation of motion of the background Higgs field is determined as
\bea
\partial^2 \ch &=& V  [k' {1 \over 4} {\cal O}_{WZ}
+ {\cal E}_1  + {\cal E}_3 V ^2 ]
+ {\cal E}_2 V ^2
+ {\cal E}_0\,,
\label{eomhiggs}\\
{\cal E}_0 &=&- m_H^2 v_1 + {\lambda_3 v_1^2 \over 2} -{\lambda_4 \over 6} v_1^3 \nnb\\
{\cal E}_1 &=&{\lambda_4 \over 2} v_1^2 -\lambda_3 v_1 +m_H^2 \nnb\\
{\cal E}_2 &=& {\lambda_3 - \lambda_4 v_1 \over 2} \nnb\\
{\cal E}_3 &=& {\lambda_4 \over 6}\nnb\\
{\cal O}_{WZ} &=& G^2 Z\cdot Z + 2 g^2 W^+ \cdot W^-\,.
\eea
When ${\cal E}_0$ and ${\cal E}_2$ vanish and $k'=k=1$, the equation of motion of
the background Higgs field restores to that of the SM.

To guarantee the quadratic terms to have the
standard form given in Eq. (\ref{stdfesm}), the gauge fixing terms
of the quantum fields are chosen as
\bea
{\cal L}_{GF,A}&=&-{1 \over 2} (\partial\cdot\hA  - i e (\hWm \cdot \bWp - \hWp \cdot \bWm))^2\cma\\
{\cal L}_{GF,Z}&=&-{1 \over 2} (\partial\cdot\hZ - {1\over 2} \sqrt{k'} G V R_2  \xi_Z
+ i {g^2 \over G} (\hWm \cdot \bWp - \hWp \cdot \bWm))^2\cma\\
{\cal L}_{GF,W}&=&- (d\cdot\hWp + {1\over 2} g \sqrt{k'} V R_2 \xi_W^+
	+ i {g^2 \over G} \bZ \cdot \hWp - i {g^2 \over G} \bWp \cdot \hZ
	+ i e \bWp \cdot \hA)\nnb\\&&(d\cdot\hWm + {1\over 2} g \sqrt{k'} V R_2 \xi_W^-
	- i {g^2 \over G} \bZ \cdot \hWm + i {g^2 \over G} \bWm \cdot \hZ
	- i e \bWm \cdot \hA)\,.
\eea
These gauge fixing terms are the 't Hooft and Feynman gauge in the backgound gauge
method, and will make the massive vector bosons, the corresponding Goldstones and
ghost, to have the same masses.

\section{The quadratic forms of the one-loop Lagrangian}
Up to one loop level, only the quadratic terms of the quantum fields
are relevant, and they can be cast into the following standard form
\bea
{\cal L}_{quad}&=&{1\over 2 } {\widehat V_{\mu}^a} \Box^{\mu\nu,ab}_{V\,V} {\widehat V_{\nu}^b}
+ {1 \over 2} \xi^i  \Box_{\xi\,\xi}^{ij} \xi^j
+ {\bar c}^a  \Box_{{\bar c}c}^{ab} c^b
+ {1\over 2} {\hat h} \Box_{hh} {\hat h}\wsep
+ {1\over 2} {\widehat V_{\mu}^{\dagger,a}} {\stackrel{\leftharpoonup}{X}}_{\xi}^{\mu,aj} \xi^j
+ {1\over 2} \xi^{\dagger,i} {\stackrel{\rightharpoonup}{X}}_{\xi}^{\nu,ib} {\widehat V_{\nu}^b}
\wsep+ {1\over 2} {\widehat V_{\mu}^{\dagger,a}} {\stackrel{\leftharpoonup}{X}}_{h}^{\mu,a} {\widehat h}
+ {1\over 2} {\widehat h} {\stackrel{\rightharpoonup}{X}}_{h}^{\mu,a} {\widehat V_{\mu}^{a}}
+ {1\over 2} \xi^{\dagger,i} X_{\xi h}^i {\widehat h}
+ {1\over 2} {\widehat h} X_{h \xi}^i \xi^{i}\,,\\
\Box^{\mu\nu,ab}_{V\, V} &=& D^{2,ab} g^{\mu\nu} + \sigma_{0,V\,V}^{ab} g^{\mu\nu} + \sigma_{2,V\,V}^{\mu\nu,ab}\,\,,\\
\Box_{\xi\,\xi}^{ij}&=&R_2 \mbra{d^{2,ij} +  \sigma_{0,\xi\xi}^{ij}
 + \sigma_{2,\xi\xi}^{ij} +  \sigma_{2, {\cal H} {\cal K},\xi\xi}^{ij}}\,,\\
\Box_{hh} &=& \partial^2 + \sigma_{h h}\,,\\
\Box_{{\bar c}c}^{ab}&=& D^{2,ab} + \sigma_{0,V \, V}^{ab}\,,\\
X_{h \xi}^i &=& X_{h \xi}^{\alpha, i} d_{\alpha}        + X_{h \xi,2}^{i}\,,\\
X_{\xi h}^i &=& X_{\xi h}^{\alpha, i} \partial_{\alpha} + X_{\xi h,2}^{i}
\label{stdfesm}
\eea
where $V^{\dagger}=(A,Z,W^{-},W^{+})$ and
$\xi^{\dagger}=(\xi_Z, \xi^{-}, \xi^{+})$.

And the covariant differential operators $D=\partial+\Gamma_V$
and $d=\partial+\Gamma_{\xi}$. The gauge connection of
vector bosons $\Gamma_V$ is defined as
\begin{displaymath}
\Gamma_{V,\mu}^{ab} =\left (\begin{array}{cccc}
0&0& i e W^{-}_{\mu}&-i e W^{+}_{\mu}\\
0&0&-i {g^2 \over G} W^{-}_{\mu} & i {g^2 \over G} W^{+}_{\mu}\\
 i e W^{+}_{\mu} & - i {g^2 \over G} W^{+}_{\mu}& - i e A_{\mu} + i {g^2 \over G} Z_{\mu} &0\\
-i e W^{-}_{\mu} &   i {g^2 \over G} W^{-}_{\mu}&0&i e A_{\mu} - i {g^2 \over G} Z_{\mu}
\end{array}\right)\,.
\end{displaymath}

The gauge connection of Goldstone bosons $\Gamma_{\xi}$ is defined as
\begin{displaymath}
\Gamma_{\xi,\mu} =\left (\begin{array}{ccc}
0&i {g\over 2} W^{-}_{\mu}&-i {g\over 2} W^{+}_{\mu}\\
i {g\over 2} W^{+}_{\mu}& - i e A_{\mu}  &0\\
- i {g\over 2} W^{-}_{\mu}&0&i e A_{\mu}
\end{array}\right)\,.
\end{displaymath}
The mass matrix of the Goldstone bosons have the following
form
\bea
\sigma_{0,\xi\xi}^{ij}=k' R_2 {1 \over 4} dia\{ G^2,
g^2,  g^2 \} V ^2\,.
\eea
Due to the gauge fixing terms we have chosen, the massive
vector bosons have the same mass with their corresponding
Goldstone bosons.

The matrix $\sigma_{2,V\,V}$ reflects that the
vector bosons are the spin $1$ particles,
and is given below as
\begin{displaymath}
\sigma_{2,V\,V }^{\mu\nu,ab} =\left (\begin{array}{cccc}
\sigma_{2,AA}^{\mu\nu}&\sigma_{2,AZ}^{\mu\nu}&\sigma_{2,AW^+}^{\mu\nu} &\sigma_{2,AW^-}^{\mu\nu} \\
\sigma_{2,ZA}^{\mu\nu}&\sigma_{2,ZZ}^{\mu\nu}&\sigma_{2,ZW^+}^{\mu\nu} &\sigma_{2,ZW^-}^{\mu\nu} \\
\sigma_{2,W^-A}^{\mu\nu} &\sigma_{2,W^-Z}^{\mu\nu}  &\sigma_{2,W^-W^+}^{\mu\nu}  &\sigma_{2,W^-W^-}^{\mu\nu} \\
\sigma_{2,W^+A}^{\mu\nu} &\sigma_{2,W^+Z}^{\mu\nu}  &\sigma_{2,W^+W^+}^{\mu\nu}  &\sigma_{2,W^+W^-}^{\mu\nu}
\end{array}\right)\,,
\end{displaymath}
and the components read
\bea
\sigma_{2,AA  }^{\mu\nu}&=&\sigma_{2,AZ}^{\mu\nu}=\sigma_{2,ZA}^{\mu\nu}=\sigma_{2,ZZ}^{\mu\nu}=0\,,\nnb\\
\sigma_{2,AW^+}^{\mu\nu}&=&-\sigma_{2,W^+A}^{\mu\nu} =   2 i e {\widetilde W^{-,\mu\nu}}\,,\nnb\\
\sigma_{2,AW^-}^{\mu\nu}&=&-\sigma_{2,W^-A}^{\mu\nu} = - 2 i e {\widetilde W^{+,\mu\nu}}\,,\nnb\\
\sigma_{2,ZW^+}^{\mu\nu}&=&-\sigma_{2,W^+Z}^{\mu\nu} = - 2 i {g^2 \over G} {\widetilde W^{-,\mu\nu}}\,,\nnb\\
\sigma_{2,ZW^-}^{\mu\nu}&=&-\sigma_{2,W^-Z}^{\mu\nu} =   2 i {g^2 \over G} {\widetilde W^{+,\mu\nu}}\,,\nnb\\
\sigma_{2,W^- W^+} &=& - \sigma_{2,W^+ W^-} = 2 i g W^{3,\mu\nu}\,,\nnb\\
\sigma_{2,W^+ W^+}&=&\sigma_{2,W^- W^-} = 0\,,
\eea
There is no deviation from the gauge theory without
spontaneous symmetry breaking if the anomalous operators $O(p^4)$
have not been taken into account.

The matrix $\sigma_{2,\xi\xi}$ indicates that
Goldstone bosons are spin $0$ particles,
and is given as
\begin{displaymath}
\sigma_{2,\xi\xi}^{ij} =\left (\begin{array}{ccc}
\sigma_{2,\xi_Z\xi_Z}& \sigma_{2,\xi_Z\xi^+} & \sigma_{2,\xi_Z\xi^-}\\
\sigma_{2,\xi^-\xi_Z}& \sigma_{2,\xi^-\xi^+} &\sigma_{2,\xi^-\xi^-}\\
\sigma_{2,\xi^+\xi_Z}& \sigma_{2,\xi^+\xi^+} &\sigma_{2,\xi^+\xi^-}
\end{array}\right)\,,
\end{displaymath}
and its components read
\bea
\sigma_{2,\xi_Z\xi_Z}&=&  {g^2 \over 2} W^{+} \cdot W^{-},\nnb\\
\sigma_{2,\xi^+\xi^+}&=&- {g^2 \over 4} W^{-} \cdot W^{-} \,,\nnb\\
\sigma_{2,\xi^-\xi^-}&=&- {g^2 \over 4} W^{+} \cdot W^{+}\,,\nnb\\
\sigma_{2,\xi_Z\xi^+}&=&\sigma_{2,\xi^+\xi_Z}={g G\over 4} W^{-} \cdot Z\,,\nnb\\
\sigma_{2,\xi_Z\xi^-}&=&\sigma_{2,\xi^-\xi_Z}={g G\over 4} W^{+} \cdot Z\,,\nnb\\
\sigma_{2,\xi^+\xi^-}&=&\sigma_{2,\xi^-\xi^+}= {g^2 \over 4}  W^{+} \cdot W^{-} + {G^2 \over 4} Z \cdot Z\,.
\eea
The matrix $\sigma_{2, {\cal H} {\cal K},\xi\xi}$ includes the terms related with the
equation of motion of the background Higgs field, which has the form
$- \sigma_{2, {\cal H} {\cal K}} {\bf 1_{3\times 3}}$, while $\sigma_{2, {\cal H} {\cal K}}$
is given as
\bea
\sigma_{2, {\cal H} {\cal K}} &=&\sigma_{2, {\cal H} {\cal K}, 0}
- {k' - 1 \over 4} {\cal O}_{WZ}
- {R_2 - 1 \over R_2} \partial^2 \ln V  \,,\\
\sigma_{2, {\cal H} {\cal K}, 0}&=&
- {1\over 4} {\cal O}_{WZ} - {\cal E}_1  - {\cal E}_3 V ^2
- {\cal E}_2 V
- {{\cal E}_0 \over V}\,,
\eea
where $\sigma_{2, {\cal H} {\cal K}, 0}$ represent the part of the SM, while
the rest of terms indicate the deviation due to the anomalous couplings in the
Higgs sector if the ${\cal E}_0$ and ${\cal E}_1$ vanish.

The $\sigma_{hh}$ is given as
\bea
\sigma_{hh} &=& - {k' \over 4} {\cal O}_{WZ}
-2 {\cal E}_2 V  - 3 {\cal E}_3 V ^2 - {\cal E}_1\,.
\eea
Due to the fact that the Higgs scalar is a singlet of
the gauge symmetry, there is no vector fields in
the $\sigma_{hh}$.

The mixing terms between the vector and Goldstone bosons
are determined as
\begin{displaymath}
{\stackrel{\leftharpoonup}{X}}_{\xi}^{\mu,aj}  =\sqrt{k'} R_2 \left (\begin{array}{ccc}
0&-i  {g^2 g' \over G}  V  W^{-,\mu} & i {g^2 g' \over G} V  W^{+,\mu}\\
 G \partial^{\mu} \ch  & i {g \over 2 G} (g^2 - g^{'2}) V  W^{-,\mu}& - i {g \over 2 G} (g^2 - g^{'2}) V  W^{+,\mu}\\
 - i  {g^2 \over 2} V  W^{+,\mu}&- g \partial^{\mu} \ch  - {1\over 2} i  g G V  Z^{\mu}&0\\
  i  {g^2 \over 2} V  W^{-,\mu}&0&- g \partial^{\mu} \ch  + {1\over 2} i  g G V  Z^{\mu}
\end{array}\right)\,,
\end{displaymath}
While the matrix ${\stackrel{\rightharpoonup}{X}}_{\xi}^{\mu,aj}$ is just
the rearrangement of the ${\stackrel{\leftharpoonup}{X}}_{\xi}^{\mu,aj}$, and
here we do not rewrite it.
The mixing terms between vector and Higgs bosons
are determined as
\bea
{\stackrel{\leftharpoonup}{X}}_{h}^{\mu,a} &=&{k'\over 2} \{0,- G^2 V  Z^{\mu},
- g^2 V  W^{+,\mu},-g^2 V  W^{-,\mu}\}\,,
\eea
The mixing terms ${\stackrel{\rightharpoonup}{X}}^{\mu,a}_h$
is the rearrangement of the ${\stackrel{\leftharpoonup}{X}}^{\mu,a}_h$.
The mixing terms between Higgs and Goldstone bosons
are determined as
\bea
X_{h \xi}^{\alpha, i}&=&\sqrt{k'} \{- G Z^{\alpha},  g W^{-,\alpha},  g W^{+,\alpha}\}\,,\\
X_{h \xi,2}^{i}      &=&\sqrt{k'} R_2  \{- {G \over 2} \partial \cdot Z,
\left[ {g \over 2} d \cdot W^- + i {1\over 2} {g'^2 \over G} W^- \cdot Z\right ],\nnb\\&&
\left [ {g \over 2} d \cdot W^+ - i {1 \over 2} {g'^2 \over G} W^+ \cdot Z \right ] \}
\eea
The terms $X_{\xi h}^{\alpha, i}$ and $X_{\xi h,0}^{i}$
are omitted here.

\section{Evaluating the traces and logarithms}
By diagonalizing the quantum fields, we can integrate the quadratic terms
of the Lagrangian by using the Gaussian integral. And the ${\cal L}_{1-loop}$
can be expressed as the traces and logarithms
\bea
S_{1-loop}&=& Tr\Box_{\bar{c}c} - {1\over 2} \mbra{ Tr\ln\Box_{V }
+Tr\ln\Box_{\xi\xi}' + Tr\ln\Box_{hh}''}
\,\,,
\label{logtrsm}
\eea
where
\bea
\Box_{\xi\xi}^{'ij}& =& \Box_{\xi\xi}^{ij} - \rvec{X}_{\xi} \Box_{V }^{-1} \lvec{X}_{\xi}\,,\\
\Box_{hh}'&=&\Box_{hh} - \rvec{X}_h \Box_{V }^{-1} \lvec{X}_h \,,\\
\Box_{hh}''&=&\Box_{hh}' - X'_{h\xi} \Box_{\xi\xi}^{'-1} X_{\xi h}\,,\\
X'_{h\xi} &=& X_{h\xi} - \rvec{X}_h \Box_{V }^{-1} \lvec{X}_{\xi}\,,\\
X'_{\xi h} &=& X_{\xi h} - \rvec{X}_{\xi} \Box_{V }^{-1} \lvec{X}_h\,,
\eea
Expanding the $Tr\ln\Box_{\xi\xi}'$ and $Tr\ln\Box_{hh}''$ with the following
relations
\bea
Tr\ln\Box_{\xi\xi}' &=& Tr\ln\Box_{\xi\xi}
+ Tr\ln( 1 - \rvec{X}_{\xi} \Box_{V }^{-1} \lvec{X}_{\xi} \Box_{\xi\xi}^{-1})\,,\\
Tr\ln\Box_{hh}''    &=& Tr\ln\Box_{hh}'
+ Tr\ln( 1 - X'_{h\xi} \Box_{\xi\xi}^{'-1} X_{\xi h}  \Box_{hh}^{'-1})\,.
\eea
Since we consider the renormalization of the theory,
so we are only interested in those
divergent terms, which can be expressed as
\bea
\int_x {\cal L}_{1-loop}&=& Tr\Box_{\bar{c}c}  - {1\over 2} \mbra{ Tr\ln\Box_{V }
+Tr\ln\Box_{\xi\xi}  + Tr\ln\Box_{hh}
\wsep - Tr(\rvec{X}_{\xi} \Box_{V }^{-1} \lvec{X}_{\xi} \Box_{\xi\xi}^{-1})
 - Tr(\rvec{X}_h \Box_{V }^{-1} \lvec{X}_h \Box_{hh}^{-1})
\wsep  - Tr(X_{h\xi} \Box_{\xi\xi}^{-1} X_{\xi h} \Box_{hh}^{-1})
\wsep - {1\over 2} Tr(X_{h\xi} \Box_{\xi\xi}^{-1} X_{\xi h} \Box_{hh}^{-1}X_{h\xi} \Box_{\xi\xi}^{-1} X_{\xi h} \Box_{hh}^{-1})
 + \cdots}
\,.
\label{divt}
\eea
Due to the property of the $Tr$, we know that the above
equation is independent of the sequence of integrating-out quantum fields.
The omitted terms are finite and will not contribute to the one-loop divergence
structures. Below we will use the Schwinger proper time and
short distance expansion method to extract the related terms from
the compact form of one loop effective Lagrangian given in Eq. (\ref{divt}).

To extract the counter terms, the following two counting
rules are necessary: one is the auxiliary counting rule,
which reads
\bea
[{\overline W_{\mu}^s}]_a=[\partial_{\mu}]_a=[D_{\mu}]_a=1\,,\,\,[v_1]_a=0\,.
\label{mcnt}
\eea
From this rule, we know
\bea
[\sigma_{2,V\,V}]_a
=[\sigma_{2,\xi\xi}]_a
=[\sigma_{2,hh}]_a
=[\sigma_{2, {\cal H} {\cal K}}]_a
=[X_{h \xi,2}^{i} ]_a
=[X_{\xi h,2}^{i} ]_a
=2\,,
\eea
\bea
[{\stackrel{\leftharpoonup}{X}}_{h}^{\mu,a}]_a
=[{\stackrel{\rightharpoonup}{X}}_{h}^{\mu,a}]_a
=[{\stackrel{\leftharpoonup}{X}}_{\xi}^{\mu,aj}]_a
=[{\stackrel{\rightharpoonup}{X}}_{\xi}^{\mu,aj}]_a
=[X_{\xi h}^{\alpha, i}]_a
=[X_{h \xi}^{\alpha, i}]_a
=1\,.
\eea
This rule is intended to count the dimensions of operators with
background fields and their differentials.

The other one is the divergence counting rule, which reads
\bea
[z^{\mu}]_d=1\,\,,\,\,[\tau]_d=-2\,.
\label{dcnt}
\eea
By using the Schwinger proper time {htkl} and short distance expansion method \cite{wkb},
after integrating over mediate coordinate spaces and proper times,
we can exact all the related divergence structures yielded by the
radiative correction. In the Schwinger proper time method, the standard propagators
can be expressed
as
\bea
\langle x|\Box^{-1,ab}_{V\,V;\mu\nu}|y\rangle=
\int_0^{\infty} \frac{d \tau}{(4 \pi \tau)^{d\over 2}}  \exp\{-\epsilon_F \tau \} \exp\left( - {z^2\over 4 \tau}\right) H^{\mu\nu,ab}_{V\,V}(x,y;\tau)\,\,,\label{vpro}\\
\langle x|\Box'^{-1,ab}_{\xi\,\xi}|y \rangle  =
\int_0^{\infty} \frac{d \tau}{(4 \pi \tau)^{d \over 2}}  \exp\{-\epsilon_F \tau \} \exp\left( - {z^2\over 4 \tau}\right) H_{\xi\xi,ab}(x,y;\tau)\,\,,
\label{spro}
\eea
where the $\epsilon_F$ is the Feynman prescription which will be taken to be
zero in the final step, and the $z$ is the distance of two event
points $y$ and $x$ which satisfies $z=y-x$.
The integral over the proper time $\tau$ and the factor
${1/(4 \pi \tau)^{d \over 2}} \exp\left( - {z^2/(4 \tau)}\right)$
conspire to separate the divergent part of the propagator.
And the $H(x,y;\tau)$ is analytic with reference to the arguments $z$ and
$\tau$, which means that $H(x,y;\tau)$ can be analytically
expanded with reference to both $z$ and $\tau$. Then we have
\bea
H(x,y;\tau)&=&H_0(x,y) +H_1(x,y) \tau + H_2(x,y) \tau^2 + \cdots\,,\\
H_i(x,y) &=& H_i(x,y)|_{x=y} + z^{\alpha} \partial_{\alpha} H_i(x,y)|_{x=y}
+ {1\over 2} z^{\alpha} z^{\beta} \partial_{\alpha}\partial_{\beta} H_i(x,y)|_{x=y}
+ \cdots
\eea
where $H_0(x,y)$, $H_1(x,y)$, and, $H_2(x,y)$ are
the Silly-De Witt coefficients. The coefficient
$H_0(x,y)$ is the Wilson phase factor, which indicates the
phase change of a quantum state from the point $x$ to
the point $y$ and reads
\bea
H_0(x,y)= \exp\int_x^y \Gamma(z)\cdot dz,
\eea
where $\Gamma(z)$ is the affine connection defined on the
coordinate point $z$.
Higher order coefficients are determined by the
lower ones by the following recurrence relation
\bea
( 1+ n + z^{\mu} D_{\mu,x} ) H_{n+1}(x,y) + (D_x^2 + \sigma) H_{n}(x,y) =0\,.
\eea
All these Silly-De Witt coefficients are gauge covariant with respect to the
gauge transformation.

For the case in the following section, here we consider the case $Tr\ln(1 - \rvec{X} \Box_{vec}^{-1} \lvec{X} \Box_{scal}^{-1})$
with $\lvec{X}= {\lvec X}_B^{\alpha} D_{\alpha} + \lvec{X}_C$ and
$\rvec{X}= {\rvec X}_B^{\alpha} D_{\alpha} + \rvec{X}_C$. The ${\lvec X}_B^{\alpha}$,
${\rvec X}_B^{\alpha}$, $\lvec{X}_C$, and $\rvec{X}_C$ have the following dimensions
in the auxiliary counting rule,
\bea
[{\lvec X}_B^{\alpha}] =[{\rvec X}_B^{\alpha}] =1\,,\,\,[\lvec{X}_C]=[\rvec{X}_C]=2\,.
\eea
By using the dimensional regularization and
the modified minimal subtraction scheme, we can extract the divergences of
the general term $Tr\ln(1 - \rvec{X} \Box_{vec}^{-1} \lvec{X} \Box_{scal}^{-1})$
which have the following forms
\bea
Tr\ln(1 - \rvec{X} \Box_{vec}^{-1} \lvec{X} \Box_{scal}^{-1}) &=& - {1 \over {\bar \epsilon}}
[t_{BB} + t_{BC} + t_{CC} + t_{BBBB}]\,,\nnb\\
t_{BB}&=&-{g_{\alpha \delta \delta' \alpha'} \over 12} \rvec{X}_B^{\alpha} D^{\delta} D^{\delta'} \lvec{X}_B^{\alpha'}
-{1 \over 4} \rvec{X}_B^{\alpha} \Gamma_{\alpha'\alpha}^{vec} \lvec{X}_B^{\alpha'}
-{1 \over 4} \rvec{X}_B^{\alpha} \lvec{X}_B^{\alpha'} \Gamma_{\alpha\alpha'}^{scal}
\wsep -{1 \over 4} \rvec{X}_B^{\alpha} H_{vec,1} \lvec{X}_B^{\alpha'}
-{1 \over 4} \rvec{X}_B^{\alpha} \lvec{X}_B^{\alpha'} H_{scal,1}
\nnb\\
t_{BC}&=&{1 \over 2} [\rvec{X}_B^{\alpha} D_{\alpha} \lvec{X}_C - \rvec{X}_C D_{\alpha} \lvec{X}_B^{\alpha} ] \nnb\\
t_{CC}&=&\rvec{X}_C \lvec{X}_C \nnb\\
t_{BBBB}&=&{g_{\alpha \alpha' \beta \beta'} \over 24} \rvec{X}_B^{\alpha} \lvec{X}_B^{\alpha'} \rvec{X}_B^{\beta} \lvec{X}_B^{\beta'}\,.
\eea
where $H_{vec,a}$ and $H_{scal,1}$ are the second Silly-De Witt coefficients, and
they might have their Lorentz indices, so the Lorentz indices should be understood
as being contracted (by multiplying metric tensors). The covariant differential operators are defined as
\bea
{\stackrel{\rightharpoonup}{X}} D {\stackrel{\leftharpoonup}{X}} &=& {\stackrel{\rightharpoonup}{X}} \partial {\stackrel{\leftharpoonup}{X}}
+ {\stackrel{\rightharpoonup}{X}} \Gamma_W {\stackrel{\leftharpoonup}{X}}
- {\stackrel{\rightharpoonup}{X}} {\stackrel{\leftharpoonup}{X}} \Gamma_{\xi}\,,\nnb\\
{\stackrel{\rightharpoonup}{X}} D D {\stackrel{\leftharpoonup}{X}} &=& {\stackrel{\rightharpoonup}{X}} \partial \partial {\stackrel{\leftharpoonup}{X}}
+ {\stackrel{\rightharpoonup}{X}} \Gamma_W \Gamma_W {\stackrel{\leftharpoonup}{X}}
+ {\stackrel{\rightharpoonup}{X}} {\stackrel{\leftharpoonup}{X}} \Gamma_{\xi} \Gamma_{\xi}
-2 {\stackrel{\rightharpoonup}{X}} \Gamma_W {\stackrel{\leftharpoonup}{X}} \Gamma_{\xi}
\nnb\\&&+2 {\stackrel{\rightharpoonup}{X}} \Gamma_W \partial {\stackrel{\leftharpoonup}{X}}
-2 {\stackrel{\rightharpoonup}{X}} \partial {\stackrel{\leftharpoonup}{X}} \Gamma_{\xi}
+  {\stackrel{\rightharpoonup}{X}} \partial \Gamma_W {\stackrel{\leftharpoonup}{X}}
-  {\stackrel{\rightharpoonup}{X}} {\stackrel{\leftharpoonup}{X}} \partial \Gamma_{\xi}\,.
\label{eqs2}
\eea
And the high rank tensor $g_{\alpha \alpha' \beta \beta'}$ is defined as
\bea
g_{\alpha\beta\gamma\delta} &=& g_{\alpha\beta} g_{\gamma\delta} + g_{\alpha\gamma} g_{\beta\delta} + g_{\alpha\delta} g_{\beta\gamma}\,.
\eea
It is symmetric under the change of its Lorentz indices.

\section{Counter terms}
By using the heat kernel method, the variant of the Schwinger proper time method,
the determinant of the D'Alamber operators can be
easily evaluated and the related divergences can
be quickly extracted out.
The divergences of the $Tr\ln\Box$ in the Eq. (\ref{divt}) are given as
\bea
{1 \over 2} {\bar \epsilon} Tr\ln\Box_{V }&=&- {20 \over 3} H_1 -{(2 g^2 + G^2) \over 16} V ^4 k'^2 R_2^2 \,,\\
{1 \over 2} {\bar \epsilon} Tr\ln\Box_{\xi\xi}&=&
+ {g^2 \over 12} H_1 + {g^{'2} \over 12} H_2 + {1\over 12} {\cal L}_1 - {1 \over 24} {\cal L}_2
 - {1 \over 24} {\cal L}_3
\wsep-{1 \over 12} {\cal L}_4 + {1 \over 48} {\cal L}_5
 - k' R_2 {G^2 g^{'2} \over 32} V ^2  Z \cdot Z
\wsep + k' R_2 {g^2 + G^2 \over 32 } V ^2  {\cal O}_{WZ}
 - k'^2 R_2^2 {2 g^2 + G^2 \over 64} V ^4
\wsep- {3 \over 4} \sigma_{2,{\cal H} {\cal K}}^2
+ \sigma_{2,{\cal H} {\cal K}} \left[ k' R_2 {2 g^2 + G^2 \over 8} V ^2 - {{\cal O}_{WZ} \over 4 } \right]
\,,\\
{1\over 2} {\bar \epsilon} Tr\ln\Box_{hh}&=&
-{k'^2 \over 16} L_5
- {k' \over 16} [\lambda_4 V ^2 + 4 {\cal E}_2 V  + 2 {\cal E}_1] {\cal O}_{WZ}
\wsep -{\lambda_4^2 \over 16} V^4 - {\lambda_4 \over 2} {\cal E}_2 V^3
- {4 {\cal E}_2^2 + {\cal E}_0 \lambda_4 \over 4} V ^2 -{{\cal E}_0 {\cal E}_1 } V
-{{\cal E}_1^2 \over 4}\,,\\
- {\bar \epsilon} Tr\ln\Box_{{\bar c}c}= &=& -{2 \over 3} H_1 + {G^2 + 2g^2 \over 32 } k'^2 R_2^2 V ^4\,.
\eea
The divergence terms from the mixing terms need some labor. But to conduct the
calculation in the coordinate space can simplify the calculation to a considerable
degree. With the formula given in the last section, contributions of the mixing terms with two
propagators can be easily extracted as
\bea
-{\bar \epsilon \over 2} Tr(\rvec{X}_{\xi} \Box_{V\, V}^{-1} \lvec{X}_{\xi} \Box_{\xi\xi}^{-1})&=&
{G^2 g^{'2} \over 8} (v_2^2 + k' V^2) Z\cdot Z V ^2
 - {g^2 + G^2 \over 8} (v_2^2 + k' V^2) {\cal O}_{WZ}
\wsep - k' {1\over 2 R_2} (2 g^2 +G^2 ) \partial \ch \cdot \partial \ch\,,\\
- {\bar \epsilon \over 2} Tr(\rvec{X}_h \Box_{V }^{-1} \lvec{X}_h \Box_{hh}^{-1})&=&
- k'^2 {g^2 g^{'2} \over 8} Z\cdot Z V ^2 - k'^2 {g^2 \over 8} V ^2 {\cal O}_{WZ}\,,\\
- {\bar \epsilon \over 2} Tr(X_{h\xi} \Box_{\xi\xi}^{-1} X_{\xi h} \Box_{hh}^{-1})&=&- {1\over 2} (t_{BB1} + t_{BB2} + t_{BC1} + t_{BC2} + t_{CC})\,,\\
t_{BB1}&=&{k'\over R_2} \bbra{- {g^2 \over 6} H_1 - {g^{'2} \over 6} H_2 + {1\over 6} {\cal L}_1 +  {1 \over 6} {\cal L}_2 +  {1 \over 6} {\cal L}_3
+  {1 \over 6} {\cal L}_4 \wsep -  {1 \over 6} {\cal L}_5
 - {1\over 2} {g'^4 \over G^4} \mbra{ {\cal L}_6 - {\cal L}_a}}
\wsep -k' { G^2 \over 4 R_2} (\partial \cdot Z)^2 -{g^2 \over 2 R_2} (d \cdot W^{+}) (d \cdot W^{-})
\wsep - i k' {g^2 g^{'2} \over G R_2} [(d \cdot W^{+}) (W^{-} \cdot Z) - (d \cdot W^{-}) (W^{+} \cdot Z)]\,,\\
t_{BB2}&=&- {k' \over R_2} \bbra{{ 1\over 4} {\cal L}_2 + {1\over 4} {\cal L}_3 +  {1\over 2} {\cal L}_4
\wsep - ({1\over 2}  - {k' \over 4}) {\cal L}_5}
  - k'^2 {G^2 g^{'2} \over 16} V^2 Z\cdot Z
\wsep + [-{k' \over 16} (k' R_2 g^2 + 2 \lambda_4) V ^2
- {k' \over 4 R_2} ( 2 {\cal E}_2 V  + {\cal E}_1)] {\cal O}_{WZ}
\wsep + {k'\over 4 R_2} \sigma_{2,{\cal H} {\cal K}} {\cal O}_{WZ} \,,\\
t_{BC1}&=&
{k'\over R_2} {g'^4 \over G^4} ({\cal L}_6 -  {\cal L}_a)
 +{k' \over R_2} {G^2 \over 2} (\partial \cdot Z)^2 + g^2 (d \cdot W^{+}) (d \cdot W^{-})
\wsep - i {k' \over R_2}  {g^2 g^{'2} \over G } [(d \cdot W^{+}) (W^{-} \cdot Z) - (d \cdot W^{-}) (W^{+} \cdot Z)]
 \,,\\
t_{BC2}&=&{g'^4 \over G^4} k' R_2 (R_2 -1)^2 \sbra{{\cal L}_6 - {\cal L}_a}
\wsep + {1 \over 2} k' R_2 (R_2 -1)^2 \mbra{
G^2 (\partial \cdot Z)^2 + 2 g^2 (d \cdot W^{+}) (d \cdot W^{-})}
\wsep + i {g^2 g'^2 \over G} k' R_2 (R_2 -1)^2 \mbra{(d \cdot W^{+}) (W^{-} \cdot Z) - (d \cdot W^{-}) (W^{+} \cdot Z)}\nnb\\
t_{CC}&=&{k' (2 - R_2)} \bbra{
-{1\over 2} {g'^4 \over G^4} \mbra{{\cal L}_6 - {\cal L}_a}
\wsep - {G^2 \over 4} (\partial \cdot Z)^2 -{g^2 \over 2} (d \cdot W^{+}) (d \cdot W^{-})
\wsep + i {g^2 g^{'2} \over 2 G}[(d \cdot W^{+}) (W^{-} \cdot Z) - (d \cdot W^{-}) (W^{+} \cdot Z)]}
\,.
\eea
The divergences of the four propagators term is given as
\bea
- {\bar \epsilon \over 4} Tr(X_{h\xi} \Box_{\xi\xi}^{-1} X_{\xi h} \Box_{hh}^{-1}X_{h\xi} \Box_{\xi\xi}^{-1} X_{\xi h} \Box_{hh}^{-1})
&=&- {k'^2\over 24 R_2^2} \sbra{2 \cal L}_4 + {\cal L}_5\,,
\eea
The sum over all contributions yields the following total divergence structures as
\bea
{\bar \epsilon} D_{tot} &=& {\bar \epsilon} (D_{tot}^{KR} +D_{tot}^{AN})\,\\
D_{tot}^{KR} &=& D_{{\widetilde \sigma}}^{KR}+D_{\sigma}^{KR}\nnb\\
D_{{\widetilde \sigma}}^{KR}&=&-{ 43 \over 6} g^2 H_1 + {1 \over 6} g^{'2} H_2
\wsep -3 {2 g^2 + G^2 \over 32} V ^2 {\cal O}_{WZ}
 -{2 g^2 + G^2 \over 2} \partial \ch \cdot \partial \ch
- {{\cal E}_1^2 \over 4}
\wsep - {1\over 2} (2 {\cal E}_2^2 + 3 {\cal E}_1 {\cal E}_3) V ^2
 - {3 \over 64} \left[ (2 g^2 + G^2) +  48 {\cal E}_3^2 \right] V ^4
\wsep - {3 \over 16} {\cal L}_5 - {{\cal E}_2 {\cal E}_1} V
- {1\over 2} {\cal E}_2 {\cal E}_3 V ^3\label{nosigs}\\
D_{\sigma}^{KR}&=&-{3 \over 4} \sigma_{2,{\cal H} {\cal K},0}^2 + {1 \over 8} \sigma_{2,{\cal H} {\cal K},0} \left[(2 g^2 + G^2) V ^2 - 3 {\cal O}_{WZ}\right]
\,,\\
D_{tot}^{AN}&=&D_{L}^{AN} + D_{\sigma}^{AN} + D_{Mass}^{AN} + D_{SF}^{AN} + D_{GF}^{AN} + D_{V}^{AN}\nnb\\
D_{L}^{AN}&=&({k' \over R_2} - 1) \bbra{{g^2 \over 12} H_1 +  {g'^2 \over 12} H_2
-  {1 \over 12 } {\cal L}_1 +  {1 \over 24 } {\cal L}_2
+ {1 \over 24 } {\cal L}_3 - ({k' \over R_2} - 1) {1 \over 12 } {\cal L}_4}
\wsep + {1 \over 48 R_2^2 } \mbra{k'^2 \sbra{-3 R_2^2 + 6 R_2 - 2} - 8 R_2 k' + 7 R_2^2 } {\cal L}_5
\wsep - {k' g'^4 \over 4 G^4} (R_2 -1 )^2 (2 R_2^2 + 1) \sbra{{\cal L}_6 - {\cal L}_a} \nnb\\
D_{\sigma}^{AN}&=&-{\bf {3 \over 4} (\sigma_{2,{\cal H} {\cal K}}^2 - \sigma_{2,{\cal H} {\cal K},0}^2 )
-{1\over 8} [(k'+2) \sigma_{2,{\cal H} {\cal K}} - 3 \sigma_{2,{\cal H} {\cal K},0} ] {\cal O}_{WZ} }
\wsep+{\bf {2 g^2 + G^2 \over 8} V ^2 (k' R_2 \sigma_{2,{\cal H} {\cal K}} -\sigma_{2,{\cal H} {\cal K},0} )}\nnb\\
D_{Mass}^{AN}&=& -3 {G^2 g'^2 \over 32} k' (k' - R_2 ) V ^2 Z \cdot Z
\wsep+{1\over 32 R_2} \bbra{V^2 \mbra{-3 g^2 R_2 k'^2 - k' \sbra{3 ( g^2 + G^2) R_2^2 + 2 \lambda_4 (R_2 -1)}
\wsep + 3 R_2 (2 g^2 + G^2)}
+ V 8 k' {\cal E}_1 (R_2 -1)
+ 4 k' {\cal E}_0 (R_2 -1)} {\cal O}_{WZ}\nnb\\
D_{SF}^{AN} &=&
i {g^2 g'^2 \over G R_2} k' (R_2 -1)^2 ( 2 R_2^2 + 1 ) [(d \cdot W^{+}) (W^{-} \cdot Z) - (d \cdot W^{-}) (W^{+} \cdot Z)]\nnb\\
D_{GF}^{AN}&=&
 - {k' \over 8 R_2} (R_2 -1)^2 (2 R_2^2 + 1) [G^2 \partial\cdot Z + 2 g^2 (d \cdot W^{+}) (d \cdot W^{-})]\nnb\\
D_{V}^{AN}&=&
- {3 \over 64} ( 2 g^2 + G^2) (k'^2 R_2^2 -1) V ^4\,.
\eea
The above equations are exact, and have not expand around the total vacuum expectation
value, $V_0$.
The $D_{tot}^{KR}$ contains all terms independent of $k'$ and $R_2$, but
dependent on the anomalous couplings of the Higgs potential. When the
anomalous couplings of the Higgs potential vanish, the $D_{tot}^{KR}$
just reduce the one loop divergence structure of the SM.
The $D_{tot}^{AN}$ includes all terms with $k'$ and $R_2$ expand around units
of the SM, and it also contains some terms dependent on the anomalous
couplings of the Higgs potential.
In the following limit,
\bea R_2 \rightarrow 1,\,\, k' \rightarrow 1\,.
\eea
each terms of the $D_{tot}^{AN}$ vanishes.

By using the equation of motion of the background Higgs field,
we get
\bea
D_{\sigma}^{KR}&=&{3 \over 16} {\cal L}_5
- {2 g^2 + G^2 \over 32} V^2 {\cal O}_{WZ}
-{3 \over 4} {{\cal E}_0^2 \over V^2}
-{3 \over 2} {{\cal E}_0 {\cal E}_1 \over V}
-{3 \over 4} \sbra{{\cal E}^2_2 + 2 {\cal E}_0 {\cal E}_2}
\wsep -{1 \over 8} \sbra{{\cal E}_0 (2 g^2 + G^2) + 12 {\cal E}_1 {\cal E}_2 + 12 {\cal E}_0 {\cal E}_3} V
\wsep -{ 1\over 8} \sbra{2 {\cal E}_1 (2 g^2 + G^2) + 6 {\cal E}_2^2 + 12 {\cal E}_1 {\cal E}_3} V^2
\wsep -{1 \over 8} {\cal E}_2 \sbra{2 g^2 + G^2 + 12{\cal E}_3 } V^3
 - {1 \over 8} {\cal E}_3 \sbra{2 g^2 + G^2 + 6 {\cal E}_3} V^4\,.
\eea
The ${\cal L}_5$ in $D_{\sigma}^{KR}$ just cancel that in the Eq. (\ref{nosigs}).
When the anomalous couplings in the Higgs potential vanish,
the $D_{tot}^{KR}$ just the master equation of the SM, and no extra divergence
will appear.

The $D_{L}^{AN}$, $D_{\sigma}^{AN}$, $D_{SF}^{AN}$,
and $D_{GF}^{AN}$ contains the anomalous operators of the
vector boson sector up to $O(p^4)$. The $D_{SF}^{AN}$ and $D_{GF}^{AN}$\footnote{
here the mass eigenstates should be understood as the combination of
the Stueckelberg fields}
indicate that the anomalous operators $L_{11}$, $L_{12}$, and $L_{13}$,
should be included when there are anomalous couplings in Higgs sector,
and the equation of motion of vector bosons can not eliminate out
these terms. As pointed out in the literature \cite{bcp}, these
three anomalous operators
are just related with the $R_b$, $B^0-{\bar B^0}$ mixing, and $b \rightarrow
s$ transitions. The $D_{Mass}^{AN}$ will contribute the parameter $\rho$,
and such a fact indicates to formulate a set of complete operators, the mass
term of $Z$ bosons with the following form
\bea
{\rho_1 -1 \over 4} k' (\ch + v_1 )^2 tr[{\cal T} V_{\mu}]^2
+ {\rho_2 -1 \over 4}  v_2^2 tr[{\cal T} V_\mu]^2
\eea
must be added to the effective Lagrangian at the $O(p^2)$ order.

The $D_{\sigma}^{AN}$ is very complicated, so
here we expand it and keep only the terms up to $O(p^4)$, which read
\bea
D_{\sigma}^{AN}&=&D_{L}^{AN} + D_{WZ}^{AN} + D_{HK}^{AN} + D_{HP}^{AN}\nnb\\
D_{L}^{AN}&=& -{1 \over 16} \mbra{k^4
- 4 \sbra{2 k'+1} k^2 + 12 k'^2 + 8 k' - 9} {\cal L}_5\nnb\\
D_{WZ}^{AN}&=&[D_{WZ,0}^{AN} +D_{WZ,1}^{AN} +D_{WZ,2}^{AN}] {\cal O}_{WZ} \nnb\\
D_{WZ,0}^{AN}&=&V_0^2 {(2g^2 +G^2) (k^2 (k'^2 +1) - 2 k'^3) \over 32 k'^2}\nnb\\
D_{WZ,1}^{AN}&=&2 V_0 \ch k \mbra{{(2g^2 + G^2) ( 1 -k'^2) \over 32 k'}
 +{m_H^2 \over 8 k^2 V_0^2} \sbra{(k'-1) k^4 \wsep + (5 k' +1) k^2 - 6 k' (k'^2 + k'-1)}}\nnb\\
D_{WZ,2}^{AN}&=& k' \ch^2 \mbra{
-{(2 g^2 + G^2) (k'^2 -1) \over 32 k'}
+{\lambda_3 \over 8 k k' V_0} \sbra{
(k'-1) k^4 + ( 5 k' +1) k^2 \wsep - 6 k' ( k'^2 + k' -1)
}
+{m_H^2 \over 4 k^2 k' V_0^2} \sbra{-4 (k' -1) k^6
\wsep + (3 k'^2 - 13 k' -2) k^4  + k' ( 5 k'+1) k^2
+ 6 k'^2 (k'^2 + k' -1) }
}\nnb\\
D_{HK}^{AN}&=&
{(2g^2 + G^2 )(k^2 -k')(2 k^2 - k') \over 4 k^2}\partial \ch \cdot\partial \ch\nnb\\
D_{HP}^{AN}&=&D_{HP,1}^{AN} +D_{HP,2}^{AN} +D_{HP,3}^{AN} +D_{HP,4}^{AN}\nnb\\
D_{HP,1}^{AN}&=&{( 2 g^2 + G^2) \sbra{k^2 (k' + 1)  - 2 k'^3} \over 8 k k'} m_H^2 \lambda_3 \ch\nnb\\
D_{HP,2}^{AN}&=& \bbra{
\mbra{k^4 (4 k' -3) - 4 k' k^2 ( 2 k'^2 - 3k' -2 )
\wsep  + k'^2 (-12 k'^2 - 16 k' + 15)} {m_H^4 \over 4 k^2 V_0^2}
\wsep +{2 g^2 + G^2 \over 8 k^2} \sbra{2 k'^3 + k^2 (-4 k'^2 + k' +1)} m_H^2
\wsep + {2 g^2 +G^2 \over 16 k k'} \sbra{k^2 ( k' +1) - 2 k'^3 } V_0 \lambda_3
}\ch^2\nnb\\
D_{HP,3}^{AN}&=& \bbra {
{2 g^2 +G^2 \over 48 k k'} \mbra{k^2 (k'+1) - 2 k'^3} \lambda_4 V_0
 +\mbra{{m_H^4 \over 2 k^3 V_0^3} \sbra{ k^6 (6 -8 k')
\wsep+ k' k^4 (8 k'^2 - 8 k' -11)
+ k'^3 ( 12 k'^2 + 16 k' -15) }
\wsep + {(2 g^2 + G^2) m_H^2 \over 4 k^3 V_0} \sbra{(k^2 - k') k'}
}
\wsep + {(2 g^2 + G^2)\lambda_3 \over 16 k^2} \mbra{2 k'^3
+ k^2 ( - 4 k'^2 + k' + 1)}
\wsep +{\lambda_3 m_H^2 \over 4 k^2 V_0^2} \mbra{ k^4  ( 4 k' - 3)- 4 k' k^2( 2 k'^2 - 3 k' -2)
\wsep + k'^2 (-12 k'^2 - 16 k' + 15) }
} \ch^3\nnb\\
D_{HP,4}^{AN}&=& \bbra{
{2 g^2 + G^2 \over 48 k^2} \mbra{2 k'^3 + k^2 (-4 k'^2 + k' +1)} \lambda_4
\wsep +\mbra{
{m_H^4 \over 4 k^4 V_0^4} \sbra{12 k^8 (4 k' -3)
+ 2 k' k^6 ( -16 k'^2 + 4 k' +31)
\wsep + k'^2 k^4 (8 k'^2 - 8 k' -11)
- 3 k'^4 ( 12 k'^2 + 16 k' -15)}
\wsep +{m_H^2 \over 12 k^4 V_0^2} \sbra{
k'^4 (2 g^2 + G^2) (k' -k^2)
+ k^2 \lambda_4 [k^4 (4 k' - 3)
\wsep -4 k' k^2 (2 k'^2 -3 k' -2)
+k'^2 ( -12 k'^2 - 16 k' +15)]}
}
\wsep +\mbra{
{\lambda_3^2 \over 16 k^2 V_0^2} \sbra{
k^4 ( 4 k' -3) - 4 k' k^2 ( 2 k'^2 - 3 k' -2)
\wsep + k'^2 (-12 k'^2 - 16 k' + 15) }
+{(2 g^2 + G^2) \lambda_3 \over 8 k^3 V_0} \sbra{k'^3 ( k^2 -k')}
}
\wsep + {\lambda_3 m_H^2 \over 2 k^3 V_0^3} \mbra{
k^6 (6 -8 k') + k^4 k' ( 8 k'^2 - 8 k' - 11)
\wsep + k'^3 (12 k'^2 + 16 k' -15)
}
}\ch^4
+ \cdots\,.
\eea
In this step, we have expanded $1/V$ and $R_2$ and one of the important
expansions is provided below
\bea
{k' \over R_2} -1 &=& (k^2 -1)
- 2 k (k^2 -k') {\ch \over V_0}
+ (k^2 - k') ( 4 k^2 - k') {\ch^2 \over 2 V_0^2}
+\cdots\,.
\eea
The $D_{WZ,0}^{AN}$, $D_{WZ,1}^{AN}$ and
$D_{WZ,2}^{AN}$ determine the renormalization constants
of the vacuum expectation value $V_0$, the anomalous
couplings $k$ and $k'$, respectively. One feature is
remarkable, that the mass of the Higgs scalar
contributes to the renormalization constants of these
two anomalous couplings.

The $D_{HP,i}^{AN}$ are related with the renormalization of
the Higgs potential. Again, the mass of
the Higgs scalar contributes to both the triple and quartic
couplings of Higgs sector.

As careful readers have noted, we have not specify the
dimension of the background Higgs field in our auxiliary counting
rule. Due to the fact that the vacuum and the background Higgs
field are originated from the same source, it is difficult to
determine a consistent Higgs field's dimension, the traditional momentum
counting rule is also invalid in this case. Here after expanding
the $R_2$, we omit those operators proportional to ${\ch}/V_0$ and
only keep those operators with mass dimension 4---
{\it i.e.} operators up to the marginal operators in the Wilsonian
renormalization framework \cite{wilson}.

Expanding $1/V$ around $R_2$ with respect to $\ch/V_0$,
some terms in $D_{\sigma}^{KR}$ and all terms of
$D_{\sigma}^{AN}$ will induce infinite divergence tower with
infinite $\ch$. Like those in the bosonic sector $O(p^4)$ \cite{our1},
the anomalous couplings of the Higgs sector up to $O^(p^2)$ also induce
infinite divergence structures, even though
at the one loop level. Such a fact is not a surprise, as
these anomalous couplings violate the requirement of the
renormalizability of the theory. However, as the
basic idea of effective theory prescribed \cite{wein},
these operators still respect the symmetry of the dynamic system.
If we include all possible operators permitted by the
specified symmetries (Lorentz symmetry and gauge symmetries),
these infinite operators still might form a set of complete operators, and
can be regarded as renormalizable, according to \cite{wein2}, though
infinite renormalization conditions are needed in order to determine
the infinite couplings of the theory, though the prediction
power of the theory is as good as the ordinary
renormalizable with finite number of couplings.

\section{Discussions and conclusions}

We have studied the one loop divergence structures of the
effective Higgs sector up to $O(p^2)$. Some features
of the divergences of this unrenormalizable theory are revealed.
We have found that, up to $O(p^4)$, the mass term of
$Z$ boson and the whole EWCL
of bosonic sector are necessary to define
a set of complete operators at the $O(p^4)$ order,
even those terms which might be eliminated by
the equation of motion of vector bosons in the EWCL $O(p^4)$
should be included.

We will study the renormalization and
provide the complete one loop renormalization
group equations of
the extended EWCL with effective
Higgs sector up to $O(p^4)$ \cite{our2}.

\acknowledgements
The work of Q. S. Yan is supported by
the Chinese Postdoctoral Science Foundation
and the CAS K. C. Wong Postdoctoral Research Award Foundation.
The work of D. S. Du is supported by the
National Natural Science Foundation of China.

\end{document}